\documentclass[pra,superscriptaddress,twocolumn,nofootinbib]{revtex4-2}

\usepackage{enumerate}
\usepackage{amsfonts,amssymb,amsmath}
\usepackage[]{graphics,graphicx,epsfig}
\usepackage{amsthm}
\usepackage{amssymb}
\usepackage{mathbbol}
\usepackage{enumitem}
\usepackage{graphicx}
\usepackage{dcolumn}
\usepackage{natbib}
\usepackage{color}
\usepackage{multirow}
\usepackage{ulem}
\usepackage{bm}
\usepackage{url}
\usepackage{float}

\usepackage[colorlinks = true, citecolor=blue, linkcolor=blue, urlcolor=blue]{hyperref}
\newcommand*{\fullref}[1]{\hyperref[{#1}]{\autoref*{#1} \nameref*{#1}}}

\def\braket#1{\left\langle#1\right\rangle}

\begin{document}

\title{Operational approach to bilocality with joint probability distributions}

\author{Kelvin Onggadinata}
\affiliation{Centre for Quantum Technologies, National University of Singapore, 3 Science Drive 2, 117543 Singapore, Singapore}
\affiliation{Department of Physics, National University of Singapore, 3 Science Drive 2, 117543 Singapore, Singapore}

\author{Pawe{\l} Kurzy{\'n}ski}
\affiliation{Institute of Spintronics and Quantum Information, Faculty of Physics, Adam Mickiewicz University, 61-614 Pozna\'n, Poland}
\affiliation{Centre for Quantum Technologies, National University of Singapore, 3 Science Drive 2, 117543 Singapore, Singapore}

\author{Dagomir Kaszlikowski}
\affiliation{Centre for Quantum Technologies, National University of Singapore, 3 Science Drive 2, 117543 Singapore, Singapore}
\affiliation{Department of Physics, National University of Singapore, 3 Science Drive 2, 117543 Singapore, Singapore}

\date{\today}

\begin{abstract}

We show an operational approach to bilocality with quasi-probability distributions and quasi-stochastic processes. This approach clearly demonstrates that negative probabilities are necessary to violate bilocality. It also highlights a subtle interplay between bilocal and local correlations and it can be easily extended to study N-locality. 

\end{abstract}

\maketitle

\section{Introduction}

Bilocality, discovered in \cite{branciard2010characterizing}, is a special case of local hidden variables (LHV) model where three observers, Alice, Bob and Charlie share two sources distributing LHVs (Alice-Bob and Bob-Charlie), instead of a common source, feeding three of them at the same time. This scenario is relevant for the fundamental research into non-local correlations, both quantum and post-quantum (for instance, PR-boxes \cite{popescu1994quantum}), but it also has a more pragmatic side related to quantum networks \cite{rosset2016nonlinear,renou2019genuine, hansenne2022symmetries} where $N$ observers share non-local correlations ($N$-locality), distributed in a variety of ways. Although, $N$-locality is a subset of LHV hypothesis that has been around since 1935 \cite{einstein1935quantum} it is a highly non-trivial one because of its non-linearity.

Let us now describe bilocality using mathematical language, most commonly used in the literature on the topic (see \cite{brunner2014bell}). In the subsequent sections we will change this language, gaining new techniques and new qualitative insights into the problem at hand. 
  
Consider three parties, Alice, Bob, and Charlie performing measurements with random inputs $x$, $y$, and $z$ and the corresponding outcomes $a$, $b$, and $c$. A tripartite collection of probability distributions is called \textit{local} if the probability of observing the outcomes $a,b,c$ given the inputs $x,y,z$ can be written as
\begin{equation}\label{eq: basic}
    P(a,b,c|x,y,z) = \sum_{\lambda} \rho(\lambda)P(a|x,\lambda)P(b|y,\lambda)P(c|z,\lambda)\,.
\end{equation}
$\lambda$'s are called local hidden variables (LHVs) and they are deterministic instructions of what output to produce for a given measurement setting. For instance, if the inputs are binary ($0,1$) and outcomes are dichotomic ($\pm 1$), a hidden variable can say: ``output $+1$ for Alice if she chooses the input $0$, $-1$ if she chooses $1$, output $-1$ for Bob if he chooses $0$ and $-1$ if he chooses $1$ and $-1$ for Charlie choosing $0$ and $+1$ if it is $1$." This can be conveniently denoted as a string $(+1,-1;-1,-1;-1,+1)$. These instructions are distributed by the source with some probability $\rho$ and, when they reach measuring apparatus, they are further locally processed with conditional probabilities $P(a|x,\lambda),P(b|y,\lambda)$ and $P(c|z,\lambda)$. 

We always insist that $P(a,b,c|x,y,z)$ is no-signalling, i.e., the distributions of local outcomes depend only on local inputs. For instance, $P(a|x)=\sum_{b,c}P(a,b,c|x,y,z)$ depends only on $x$, not on $y$ or/and $z$.

\begin{figure}[!htb]
    \centering
    \includegraphics[width=\linewidth]{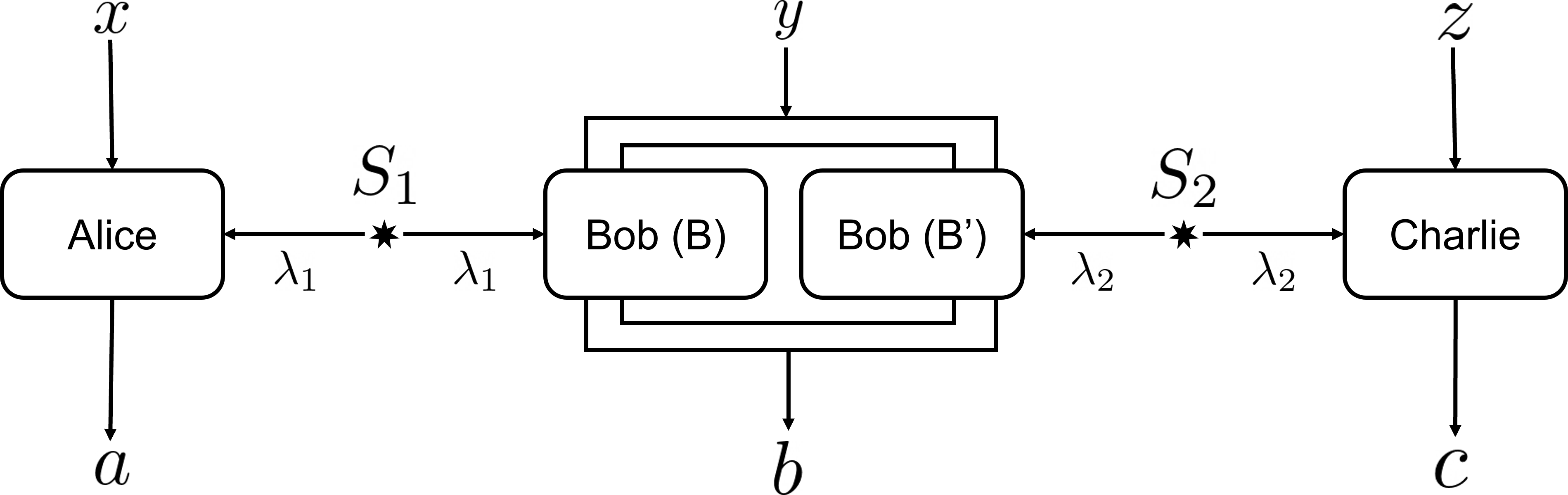}
    \caption{Illustration of a typical bilocal hidden variable model. Two independent sources, $S_1$ and $S_2$, emit hidden variables $\lambda_1$ and $\lambda_2$ to the pair Alice-Bob(B) and Bob(B')-Charlie. Alice, Bob, and Charlie perform random measurements $x,y,z$ with the corresponding outcomes $a,b,c$. Note that Bob measures jointly both subsystems he receives.}
    \label{fig: bilocal hvm}
\end{figure}

This scenario assumes that the instructions encoded in $\lambda$'s are distributed to all parties simultaneously. This can be relaxed. For instance, in Fig. \ref{fig: bilocal hvm}, there are two independent sources $S_1$ and $S_2$, a scenario we call \textit{bilocal} \cite{branciard2012bilocal}. This assumption further decomposes Eq. \eqref{eq: basic} into
\begin{eqnarray}\label{eq: bilocal hvm}
P(a,b,c|x,y,z) & = & \sum_{\lambda_1 \lambda_2} \rho_1(\lambda_1)\rho_2(\lambda_2)P(a|x,\lambda_1) \nonumber \\ & & \quad \times   P(b|y,\lambda_1,\lambda_2)P(c|z,\lambda_2)\, .
\end{eqnarray}
This relaxation has lead to new, interesting scenarios, like the one discussed in \cite{branciard2010characterizing} and has been subsequently generalised to a more complex setups \cite{tavakoli2022bell}. %

To proceed, we first introduce a notation, which we will use interchangeably with the previous one:
\begin{equation*}
    P(a,b,c|x,y,z) \leftrightarrow P(a_x,b_y,c_z)\,.
\end{equation*} 
Now, a joint probability distribution is a probability distribution $q$ over all measurement outcomes $\vec{a}=(a_0,...,a_{|X|-1})$, $\vec{b}=(b_0,...,b_{|Y|-1})$, $\vec{c}=(c_0,...,c_{|Z|-1})$ with the following properties:
\begin{eqnarray}
\sum_{\vec{a},\vec{b},\vec{c}}q(\vec{a},\vec{b},\vec{c})&=&1\label{eq: cond1}\\
\sum_{\vec{a}\setminus a_x,\vec{b}\setminus b_y,\vec{c}\setminus c_z}q(\vec{a},\vec{b},\vec{c})&=&P(a_x,b_y,c_z)\label{eq: cond2}.
\end{eqnarray} 
Here, $|X|, |Y|$ and $|Z|$ denote the number of measurement settings for Alice, Bob and Charlie, and $\vec{a}\setminus a_x$ indicates all the variables in $\vec{a}$, excluding $a_x$. In this paper we consider positive joint probability distributions (JPD), i.e., all probabilities are positive, and negative joint quasi-probability distributions (JQD) where some probabilities can be negative. Note that both JPD and JQD are automatically no-signalling. 

Fine \cite{fine1982hidden} showed that one can find a JPD with properties \eqref{eq: cond1} and \eqref{eq: cond2} if and only if the marginal distribution $P$ in \eqref{eq: cond2} is local. If $P$ is nonlocal, some of $q$'s entries must be negative. Of course, \eqref{eq: cond2} guarantees that all observable statistics are semi-positive or else we are forced to find an operational interpretation of negative probabilities, which goes beyond the scope of this paper.

When all outputs are binary, any joint probability distribution (JPD and JQD) can be written in a compact form by labelling the outcomes as $\{+1,-1\}$. In the simplest Bell-CHSH scenario with two parties, two inputs each, we can change $(a_0a_1b_0b_1)\to(s_0s_1s_2s_3)$ for convenience, giving us
\begin{eqnarray}
    q(s_0s_1s_2s_3) &=& \frac{1}{2^4}\Big[1 + \sum_{j=0}^3 s_j E^{(1)}_{j} + \sum_{j < k}s_js_k E^{(2)}_{jk} \nonumber \\ 
    & & \, + \sum_{j<k<l}s_js_ks_lE^{(3)}_{jkl} + s_0s_1s_2s_3 E^{(4)}\Big]
\end{eqnarray}
where $E^{(1)}_j=\langle s_j\rangle_q$, $E^{(2)}_{jk}=\langle s_js_k\rangle_q$ etc. The observable statistics drawn from $P$ determine the correlation functions (correlators) $E^{(1)}_j$'s, $E^{(2)}_{02}$, $E^{(2)}_{03}$, $E^{(2)}_{12}$ and $E^{(2)}_{13}$. The remaining, unobservable correlators, should satisfy $|E|\leq1$ to be consistent with observable outcomes $\pm 1$ but are arbitrary otherwise. The extension to more inputs and more parties is straightforward.

\section{The bilocal scenario with joint probability distributions}

\begin{figure}[!htb]
    \centering
    \includegraphics[width=\linewidth]{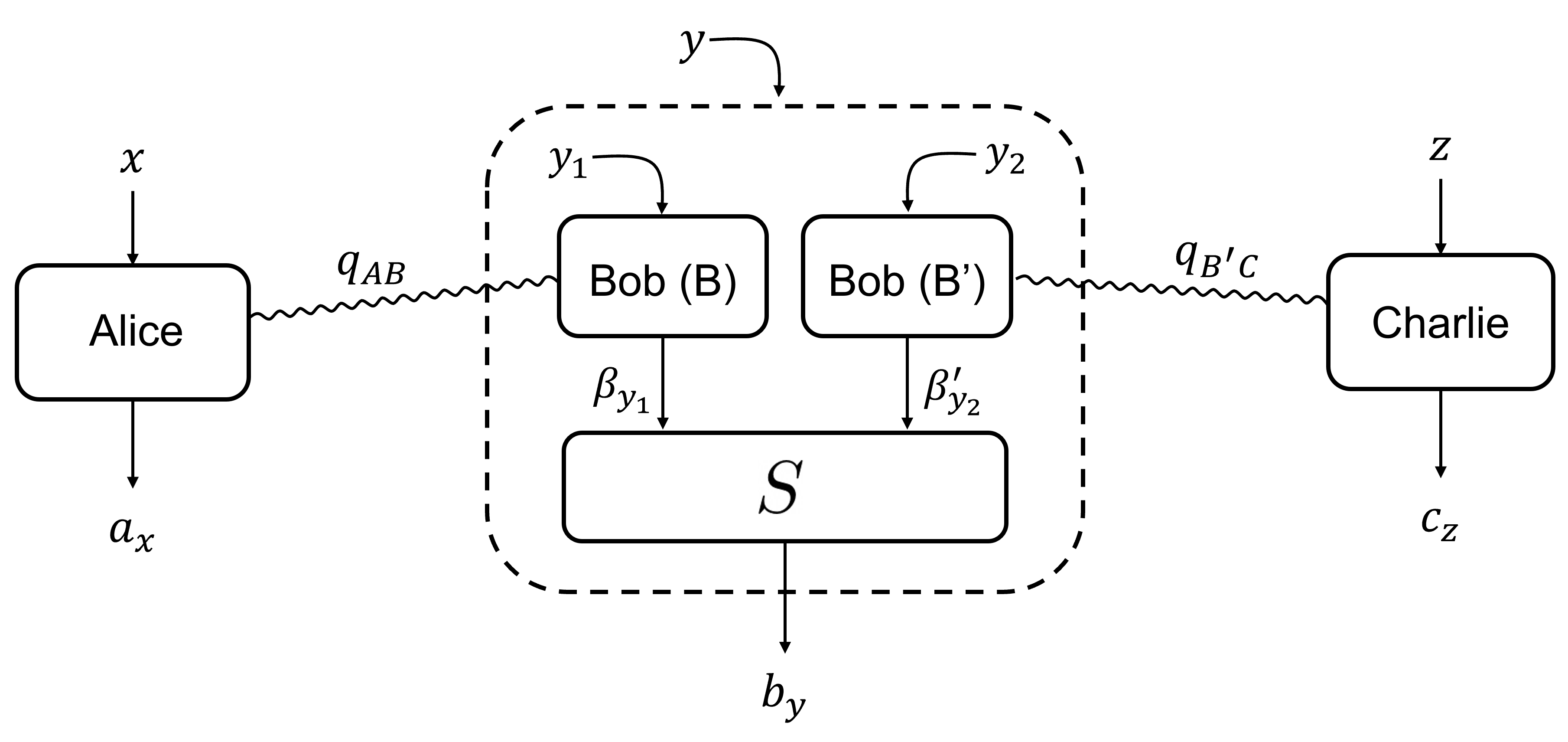}
    \caption{Alternative description of the bilocal model in terms of joint probability distributions with each boxes having binary inputs and outputs. Initially, two independent sources distribute local hidden variables with joint probability distributions $q_{AB}(a_0a_1\beta_0\beta_1)$ and $q_{B'C}(\beta_0'\beta_1'c_0c_1)$ to the pair Alice-Bob(B) and Bob(B')-Charlie, respectively. The joint measurement on Bob's part is replaced by a local processing $S$ acting on the previously obtained outcomes $\{\beta_0,\beta_1,\beta_0',\beta_1'\} \rightarrow \{b_0,b_1\}$.}
    \label{fig: bilocal jpd}
\end{figure}

We illustrate the bilocal scenario with JPDs in the Fig. \ref{fig: bilocal jpd}. It can be mathematically written as 
\begin{equation}\label{eq: bilocal jpd}
r(\vec{a},\vec{b},\vec{c}) = \sum_{\vec{\beta},\vec{\beta}'}S(\vec{b}|\vec{\beta},\vec{\beta}')r_0(\vec{a},\vec{\beta},\vec{\beta}',\vec{c})\, ,
\end{equation}
where $\vec{a} = (a_0,a_1,\dots,a_{N-1})$ is the vector of measurement outcomes $a_k=\pm 1$ ($k=0,1,\dots, N-1$) for $N$ settings (inputs). Similar definition applies to $\vec{b},\vec{c},\vec{\beta},\vec{\beta}'$ but these vectors do not necessarily have the same number of measurement settings. $S(\vec{b}|\vec{\beta},\vec{\beta}')$ represents Bob's data processing. Bilocality imposes the following constraints:
\begin{enumerate}[label=(\roman*)]
    \item {\it Independent preparation} --- the initial JPD takes the factorization form
    \begin{equation}
        r_0(\vec{a},\vec{\beta},\vec{\beta}',\vec{c}) = q_{AB}(\vec{a},\vec{\beta})q_{B'C}(\vec{\beta}',\vec{c})\, ,
    \end{equation}
    where $q_{AB}(\vec{a},\vec{\beta}),q_{B'C}(\vec{\beta}',\vec{c}) \geq 0$ are generated by two completely independent local sources.
    \item {\it Local processing} --- Mirroring the assumptions made in \eqref{eq: bilocal hvm}, Bob has a device that stochastically processes data received from both sources. This device is mathematically a stochastic process  $S(\vec{b}|\vec{\beta},\vec{\beta}')$ such that $S(\vec{b}|\vec{\beta},\vec{\beta}')\geq 0$ and $\sum_{\vec{b}}S(\vec{b}|\vec{\beta},\vec{\beta}') = 1$.  
\end{enumerate}

\subsection{Bilocal inequality}

To warm up and see the joint probability distribution formalism at work, let us re-derive the bilocal inequality from \cite{branciard2012bilocal}. In the scenario considered there, Alice and Charlie have two inputs and two dichotomic outputs, i.e., $\vec{a}=(a_0,a_1)$ and $\vec{c}=(c_0,c_1)$ with $a_x,c_z\in\{+1,-1\}$.

Let us define
\begin{eqnarray}
    \braket{I}_r &=& \frac{1}{4} \sum_{i,j=0,1}\braket{A_iB_0C_j}_r\, , \label{eq: corr func I}\\
    \braket{J}_r &=& \frac{1}{4} \sum_{i,j=0,1}(-1)^{i+j}\braket{A_iB_1C_j}_r \label{eq: corr func J}\, ,
\end{eqnarray}
where 
\begin{equation}
 \braket{A_iB_jC_k}_r = \sum_{a_0a_1}\sum_{b_0b_1}\sum_{c_0c_1}a_ib_jc_k r(a_0a_1b_0b_1c_0c_1)\, .
\end{equation}
We now show that any bilocal joint probability distribution given in \eqref{eq: bilocal jpd} together with assumptions (i)-(ii) imply
\begin{equation}\label{eq: bilocal inequality}
    \sqrt{|\braket{I}_{r}|} + \sqrt{|\braket{J}_{r}|} \leq 1\, .
\end{equation}
First, let us take the absolute value of \eqref{eq: corr func I} and use the inequality $|\sum_k x_k| \leq \sum_k |x_k|$ to obtain 
\begin{eqnarray}
|\braket{I}_{r}| & \leq & \sum_{\vec{a},\vec{b},\vec{c}} \sum_{\vec{\beta},\vec{\beta}'}\frac{1}{4}|b_0(a_0+a_1)(c_0+c_1)| \nonumber \\ 
& & \quad \times S(\vec{b}|\vec{\beta},\vec{\beta}') q_{AB}(\vec{a},\vec{\beta})q_{B'C}(\vec{\beta}',\vec{c}) \nonumber \\
& = & \sum_{\vec{a},\vec{\beta}}\frac{|a_0+a_1|}{2}q_{AB}(\vec{a},\vec{\beta}) \nonumber \\
& & \times \sum_{\vec{\beta}',\vec{c}} \frac{|c_0+c_1|}{2} q_{B'C}(\vec{\beta}',\vec{c}) \nonumber \\ 
& & \times \sum_{\vec{b}} |b_0|S(\vec{b}|\vec{\beta},\vec{\beta}')\, .
\end{eqnarray}
Since $b_0$ can only take values $\pm 1$, we have $\sum_{\vec{b}}|b_0|S(\vec{b}|\vec{\beta},\vec{\beta}') = \sum_{\vec{b}}S(\vec{b}|\vec{\beta},\vec{\beta}') = 1$. The above simplifies to 
\begin{equation}
|\braket{I}_{r}| \leq \sum_{a_0a_1}\frac{|a_0+a_1|}{2}p_A(\vec{a})\sum_{c_0c_1} \frac{|c_0+c_1|}{2} p_C(\vec{c})
\end{equation}
with $p_A(\vec{a})=\sum_{\vec{\beta}}q_{AB}(\vec{a},\vec{\beta})$, $p_C(\vec{c})=\sum_{\vec{\beta}'}q_{B'C}(\vec{\beta}',\vec{c})$.

In a similar manner we get 
\begin{equation}
|\braket{J}_{r}| \leq \sum_{a_0a_1}\frac{|a_0-a_1|}{2}p_{A}(\vec{a})\sum_{c_0c_1} \frac{|c_0-c_1|}{2} p_{C}(\vec{c})\, .
\end{equation}
We use the inequality $\sqrt{uv}+\sqrt{u'v'} \leq \sqrt{u+u'}\sqrt{v+v'}$ for $u,v,u',v'\geq 0$ and obtain
\begin{eqnarray}
    & &\sqrt{|\braket{I}_{r}|} + \sqrt{|\braket{J}_{r}|}   \\ 
    & &\quad \leq \sqrt{\sum_{a_0a_1}\left(\frac{|a_0+a_1|}{2}+\frac{|a_0-a_1|}{2}\right)p_{A}(a_0a_1)} \nonumber \\ 
    & &\quad \times \sqrt{\sum_{c_0c_1}\left(\frac{|c_0+c_1|}{2}+\frac{|c_0-c_1|}{2}\right)p_C(c_0c_1)} = 1 \, . \nonumber
\end{eqnarray}
The last equality comes from $|a_0+a_1|/2 + |a_0-a_1|/2 = \max\{|a_0|,|a_1|\} = 1$ and $|c_0+c_1|/2 + |c_0-c_1|/2 = \max\{|c_0|,|c_1|\} = 1$. The remaining summation over $p_{A}(a_0a_1)$ and $p_{C}(c_0c_1)$ simply returns 1.

\section{Violations of bilocality}

In this section we use the JPD/JQD formalism to explain how violation of bilocality reported in \cite{branciard2012bilocal} happens in simple, operational terms. To be clear, we focus on this particular scenario but whatever we present here can be applied to any $N$-locality scenario with appropriate modifications. 

In \cite{branciard2012bilocal} the authors model Bob's data processing on entanglement swapping. They also assume that Alice and Bob(B), as well as, Bob(B') and Charlie, share maximally entangled states. Their measurement settings are carefully chosen to avoid revealing any quantum non-locality. These assumptions can be now directly mapped into our formalism: (1) we need to choose Alice-Bob(B) and Bob(B')-Charlie joint probability distributions and (2) Bob(BB') data processing $S$. 

What follows is an extension of the joint probability distribution approach considered in \cite{abramsky2014operational}, now applied to bilocality instead of just locality. Locality is equivalent to JPD and nonlocality requires JQD. On the other hand, bilocal correlations are included in local correlations, therefore there are some local correlations that are not bilocal and should violate the bilocality inequality. Such correlations must be described by a JPD, hence non-bilocality must be manifested by some more subtle properties of JPDs. Below we find what these properties are.
  
To this end, consider the following joint probability distribution
\begin{subequations}
\begin{align}
    q_{AB}(a_0a_1\beta_0\beta_1) & = \frac{1}{16}\Big[1 + \mu_1(a_0\beta_0+a_0\beta_1 + a_1\beta_0-a_1\beta_1)\Big] \\
    q_{B'C}(\beta_0'\beta_1'c_0c_1) & = \frac{1}{16}\Big[1 + \mu_2(\beta_0'c_0+\beta_0'c_1 + \beta_1'c_0-\beta_1'c_1)\Big]
\end{align}
\end{subequations}
where $\mu_{1,2}\in [0,1]$. The above $q$'s have the following physical interpretations: (i) for $0\leq \mu_{1,2} \leq \frac{1}{2}$ we have JPDs that saturate the LHV bound for the Bell-CHSH inequality; (ii) for $\frac{1}{2}<\mu_{1,2} \leq \frac{1}{\sqrt{2}}$ we have JQDs corresponding to the quantum mechanical violations of the Bell-CHSH inequality; (iii) for $\frac{1}{2} < \mu_{1,2} \leq 1$ we deal with JQDs representing super quantum correlations generated by noisy PR-boxes; (iv) the region for $\mu_{1,2}>1$ has no known physical meaning because one would observe negative probabilities in observable probability marginals.

Next we chose Bob's stochastic process as
\begin{eqnarray}\label{eq: coupler}
    & & S_\eta(b_0b_1|\beta_0\beta_1\beta_0'\beta_1') \nonumber \\
    & & \qquad = S_\eta(b_0|\beta_0\beta_0')S_\eta(b_1|\beta_1\beta_1')\nonumber \\
    & & \qquad = \frac{1}{2}\Big[1 + \eta b_0\beta_0\beta_0'\Big]\times \frac{1}{2}\Big[1 + \eta b_1\beta_1\beta_1'\Big]\, ,
\end{eqnarray}
where $S_\eta$ is positive for $\eta \leq 1$. Inserting these expressions into \eqref{eq: bilocal jpd}, we get
\begin{eqnarray}
r(a_0a_1b_0b_1c_0c_1)&=& \frac{1}{2^6}\Big[1 + \eta\mu_1\mu_2\big\{b_0(a_0+a_1)(c_0+c_1)\nonumber\\ && \qquad + b_1(a_0-a_1)(c_0-c_1)\big\}\Big]\,,
\end{eqnarray}
whence
\begin{equation}
    \sqrt{|\braket{I}_r|}+\sqrt{|\braket{J}_r|} = 2\sqrt{\eta\mu_1\mu_2}\,.
\end{equation}
Thus violations of the bilocal inequality happen when $\eta\mu_1\mu_2 >\frac{1}{4}$. 

Let us summarize the bounds on $\eta\mu_1\mu_2$ for the different regions
\begin{equation}
    \eta\mu_1\mu_2 \stackrel{\text{BL}}{\leq}\frac{1}{4} \stackrel{\text{BQ}}{\leq}\frac{1}{2} \stackrel{\text{L,BNS}}{\leq} 1
\end{equation}
where BQ and BNS refers to bi-quantum and bi-no-signaling, respectively. Note that the upper bound is imposed such that negative probabilities are not observed at the level of observable marginals $r_{\eta\mu_1\mu_2}(a_jb_kc_l)$.

\begin{figure}[!htb]
    \centering
    \includegraphics[width=0.95\linewidth]{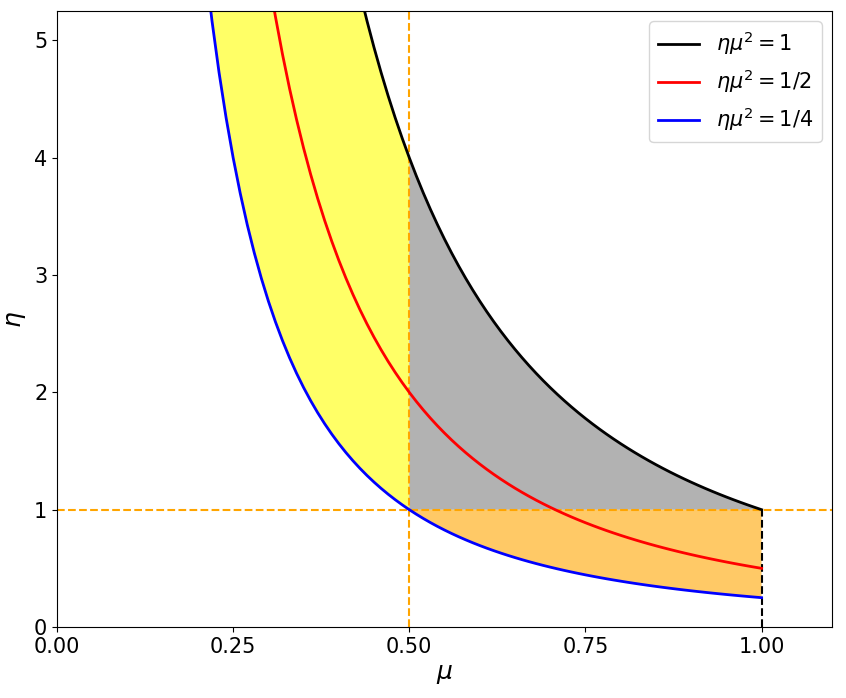}
    \caption{Strength of the nonbilocal correlations as a function of $\eta$ and $\mu$. The regions below the blue, red and black lines corresponds to bilocal, bi-quantum, and bi-no-signaling correlations, respectively. The three shaded regions correspond to different origins of negativity. The details are described in the main text.}
    \label{fig: bilocal negative region}
\end{figure}

The violation of the bilocal inequality depends on the values of $\eta$ and $\mu_{1,2}$. In Fig. \ref{fig: bilocal negative region} we show different violation regions that correspond to different origins of negativity. Here we assume that $\mu_1=\mu_2=\mu$, so $q_{AB}$ and $q_{B'C}$ are symmetric in their correlations. We group the violations as follows:
\begin{itemize}
    \item JQD and positive $S_{\eta}$ (orange shaded region): $\mu > \frac{1}{2}$ and $\eta\leq 1$. 
    \item JPD and negative $S_{\eta}$ (yellow shaded region): $\mu
    \leq\frac{1}{2}$ and $\eta>1$.
    \item JQD and negative $S_{\eta}$ (gray shaded region): $\mu > \frac{1}{2}$ and $\eta>1$.
\end{itemize}
Therefore, negative probabilities are required either in the initial joint probability distributions shared between Alice-Bob(B) and Bob(B')-Charlie, or in the local processing performed by Bob(BB'). This is also true for $\mu_1\neq \mu_2$. 

The above classification of different mechanisms responsible for violations of bilocality is one of the main results of this paper. We can now see that violations of bilocality require negative probabilities either in the initial probability distributions shared by the observers or in Bob's (BB') data processing or both. We discussed a similar mechanism in the context of quantum nonlocality in \cite{onggadinata2021local}. This brings locality and bilocality violations into a universal mathematical and conceptual framework.

Let us make some remarks and observations. Note that the JQD
\begin{eqnarray}
    r_{\text{BQ}}(a_0a_1b_0b_1c_0c_1) &=& \frac{1}{2^6}\Big[1 + \frac{1}{2}\Big\{b_0(a_0 + a_1)(c_0 + c_1) \nonumber \\ 
    & & \quad + b_1(a_0 - a_1)(c_0 - c_1)\Big\}\Big] \label{eq: biquantum jqd}
\end{eqnarray}
saturates the maximal BQ bound. If we mix $r_{\text{BQ}}$ with white noise, we get a noisy JQD
\begin{eqnarray}
    r_{\text{BQ}}^V(a_0a_1b_0b_1c_0c_1) &=& V r_{\text{BQ}}(a_0a_1b_0b_1c_0c_1) + \frac{1-V}{2^6} \nonumber \\
    & =& \frac{1}{2^6}\Big[1 + \frac{V}{2}\Big\{b_0(a_0 + a_1)(c_0 + c_1) \nonumber \\ 
    & & \quad + b_1(a_0 - a_1)(c_0 - c_1)\Big\}\Big]\, ,
\end{eqnarray}
where $V\in [0,1]$ is customarily called visibility. The above JQD becomes bilocal when $V\leq \frac{1}{2}$, in which case it also becomes a JPD. This is in agreement with \cite{branciard2012bilocal}. However, note that the above is a global noise admixture. If each source that produces the initial correlations introduces their own white noise with probability $1-v_p$ and $1-v_q$, then $V = v_pv_q$. 

In addition, $r_{\text{BQ}}$ is not unique. One could choose a different JQD that mimics $r_{\text{BQ}}$, i.e., having the same observable probabilities. In fact, there exists a JPD that is equivalent to $r_{\text{BQ}}$ and still violates the bilocal inequality. For instance, one could consider 
\begin{eqnarray}
    & & r_{\text{BQ}}^+(a_0a_1b_0b_1c_0c_1) \nonumber \\
    & & = \frac{1}{2^6}\Big[1 + \frac{1}{2}\Big\{b_0(a_0 + a_1)(c_0 + c_1) \nonumber \\
    & & \quad + b_1(a_0 - a_1)(c_0 - c_1) + a_0a_1c_0c_1\Big\}\Big]\, .
\end{eqnarray}
It is easy to check that it is nonnegative. However, it is also inseparable
\begin{equation}
    \sum_{b_0b_1}r_{\text{BQ}}^+(a_0a_1b_0b_1c_0c_1)\neq p(a_0a_1)q(c_0c_1)\, .
\end{equation}
Hence, it does not come from two independent sources, it is not interesting to us. Moreover, the JQD in Eq. \eqref{eq: biquantum jqd} is optimal in the sense that it has the minimal amount of negativity while staying separable. Here, we use the $L_1$-norm measure of negativity \cite{oas2014exploring} defined as 
\begin{equation}
    N(q) = \sum_{x}|q(x)| \geq 1
\end{equation}
with equality saturated iff $q$ is completely positive.

Interestingly, since $b_0$ and $b_1$ are co-measurable in the entanglement swapping scenario utilised in \cite{branciard2012bilocal}, we expect that the marginals $r(a_xb_0b_1c_z)$ are semi-positive. Indeed, they behave properly up to the maximal BQ bound and break down once we exceed it. This is not surprising because above this bound we deal with noisy PR-boxes that do not have a quantum mechanical counterpart. We designed Bobs (BB') (quasi)stochastic process to simulate quantum mechanical measurement, not an imaginary PR-box measurement. Perhaps this is an argument one can use to reject PR-boxes as non physical but more research is needed, which goes beyond the scope of this paper. 

Finally, let us discuss Bob's transformation $S_\eta$. When $\eta=1$, $S_{\eta=1}$ is the classical \texttt{AND} operation on the bits $(\beta_0,\beta_0') \to b_0$ and $(\beta_1,\beta_1')\to b_1$. $S_{\eta=1}$ is a positive stochastic matrix, consistent with the requirement that for any JQD $p$ and $q$ the final JQD $r$ has positive observable marginals. Note that any stochastic matrix $S_{\eta}$, acting on JQD, does not increase its negativity. When $\eta>1$, one can think of $S_\eta$ as a {\it nebit} (negative bit) controlled operation where \texttt{AND} is performed with probability $(1+\eta)/2$ and  \texttt{NAND} with probability $(1-\eta)/2$. Nebit, introduced in \cite{onggadinata2021local}, is a catalyst turning stochastic operations into quasi-stochastic ones and it is useful to quantify non-classical behaviour or in this case, non-bilocality.

\section{Final remarks}

Bilocality is a relatively new phenomenon in the topic of quantum and post-quantum nonlocality that has so far generated a host of interesting and still unsolved problems, see for instance \cite{renou2019genuine,tavakoli2022bell}. It has also practical implications with the advent of quantum internet \cite{kimble2008quantum,wehner2018quantum} and distributed quantum computing \cite{acin2006bell,acin2007device}. 

In this paper, we present a certain way to analyse bilocality and more generally, $N$-locality, based on a well established, albeit still niche, research into quasi-probabilities and quasi-stochastic processes. We demonstrated that our approach recovers known results (the bilocality inequality from \cite{branciard2012bilocal}) but it also offers operational insights (stochastic and quasi-stochastic data processing) into the mechanisms responsible for violations of bilocality that have not been yet discussed in the literature. 

Extensions to $N$-locality are straightforward as adding more independent JPDs and/or JQDs and more quasi-stochastic and/or stochastic processes does not change the formal structure of our approach.

\section*{Acknowledgements}
We thank Valerio Scarani for suggestions and discussions. This research is supported by the National Research Foundation, Singapore and A*STAR under its CQT Bridging Grant. PK is supported by the Polish National Science Centre (NCN) under the Maestro Grant no. DEC-2019/34/A/ST2/00081.

\bibliography{references}

\end{document}